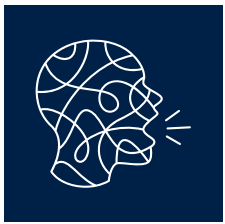
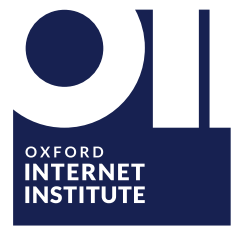
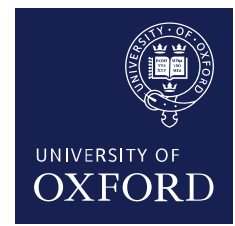

# Junk News & Information Sharing During the 2019 UK General Election

Nahema Marchal, Bence Kollanyi, Lisa-Maria Neudert, Hubert Au, Philip N. Howard

## Introduction

More than a year after the Cambridge Analytica scandal exposed rampant and unlawful harvesting of user data to influence politics, voter manipulation and election meddling are still major concerns in the United Kingdom. In a bid to sway voters in the lead up to the 2016 EU Referendum, manipulative actors deployed sophisticated digital tactics, ranging from bots to amplifier accounts and targeted ads.[1] We also know from previous research that ideologically extreme, hyper-partisan and conspiratorial content shared over Twitter and Facebook were prevalent during the 2017 General Election campaign.[2]

At a time when trust in the media and the political establishment is challenged[3], policy-makers have taken steps to address issues linked with information operations, dishonest campaigning practices and obscure political advertising.[4] In 2018, DCMS launched an "Online Harms White Paper" calling for a new regulatory framework to improve citizen safety online.[5] The ICO released a draft framework for a code of practice for the use of data in political campaigns.[6] The UK Electoral Commission also shared recommendations for increasing the transparency of political campaigns[7]. Increasingly, social media firms too are taking to self-regulation of their platforms. During the time of the 2019 UK General Election campaign alone, Google, Facebook and Twitter have implemented new ad policies and product features.[8]

Today, an estimated 75% of the British public access information about politics and public life online, and 40% do so via social media.[9] With this context in mind, we investigate information sharing patterns over social media in the lead-up to the 2019 UK General Elections, and ask: **(1)** What type of political news and information were social media users sharing on Twitter ahead of the vote? **(2)** How much of it is extremist, sensationalist, or conspiratorial junk news? **(3)** How much public engagement did these sites get on Facebook in the weeks leading up to the vote? And **(4)** What are the most common narratives and themes relayed by junk news outlets?

To answer these research questions, we collected 1.76 million tweets related to the UK General Elections using the Twitter Streaming API between 13th and 19th November. These were collected from 284,265 unique users using a list of 40 election-related hashtags associated with the primary political parties in the UK, the 2016 EU referendum, and the 2019 General Election itself. From this sample, we extracted 308,493 tweets containing a URL link, which pointed to a total of 28,532 unique URLs.

Sources that were shared ten times or more across our collection period were manually classified by a team of three coders based on a rigorous grounded typology developed and refined through the project's previous studies of nine elections in several countries around the world.[10,11] After two rounds of test coding, our team reached a Krippendorff's alpha of 0.77, indicating high inter-coder reliability. Using this technique, we were able to successfully label nearly 96.4% of all links shared in our sample (see online supplement for full specification of methods).

To better understand the nature of the Twitter conversation about UK politics during the campaign, we analysed the relative use of party and issue-based hashtags, as well as levels of high-frequency tweeting and patterns of information sharing for our sample week. We then measured and compared public engagement with producers of junk and professional news over Facebook in the first three weeks of the campaign (6th–27th November) using the CrowdTangle social monitoring tool.[11] Lastly, we conducted a thematic analysis of the junk news stories receiving the most engagement on UK Facebook during our data collection period, to shed light on the political narratives favoured by junk news outlets.

**Our main findings are:**

- Fewer than 2% of links shared on Twitter during our data collection period were identified as Junk News, a tenth of what we had found in 2017. Instead, Professional News Content constituted over 57% of total traffic.

- Labour-related hashtags topped Twitter traffic during our entire data collection period. This trend reversed on the night of the first televised leaders' debate during which traffic around Conservatives' hashtags rose three-fold.

- While professional news outlets are more prolific, and their stories are shared by far more people, posts from junk news outlets trigger more extreme reactions from Facebook users.

- The most engaging stories produced by junk news outlets and shared over Facebook during the campaign were indictments against the mainstream media, and the BBC in particular, followed by ad-hominem attacks against specific candidates.

# Trends in News and Information Sharing over Twitter

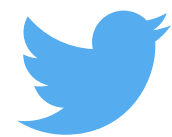

In contrast to our previous findings from the 2017 UK General Election, we found very little junk news circulating over Twitter during our data collection period (less than 2%) (see *Table 1*). While we found almost no trace of known Russian sources of propaganda, nearly half of junk outlets identified in our sample were foreign, with most of them located in the US, Germany or Canada. Articles from professional news outlets and tabloids, on the other hand, made up the great majority of Twitter traffic, representing respectively 45% and nearly 12% of all links. Within the professional news outlets shared, The *Guardian* was the most popular, followed by the BBC and *The Independent* (see online supplement for full table). This was closely followed by links to political party, expert and government websites which, taken together, made up 16% of overall traffic from 13th to 19th November. Content produced by civil society at large was also widely circulated: for every ten links shared over Twitter, one redirected to a third-section source such as a personal blog or a non-profit organization's website.

Having classified the main sources of political news and information shared over Twitter, we explored the traffic produced by party-related hashtags. Our analysis shows that the Labour Party dominated the conversation for most of the sample week, generating 20% all of tweets (see *Figure 1*). This trend reversed dramatically on the day of the first televised debate between the Prime Minister, Boris Johnson, and the Leader of the Opposition, Jeremy Corbyn.

As *Figure 1* shows, on November 19th Twitter activity around Conservatives-related hashtags increased three-fold compared to the day prior — reaching 20,680 tweets/hour at its peak. This spike coincided with the controversial decision by the Conservatives to rebrand their Twitter handle to "factcheckUK" ahead of the leaders' debate[12] Turning now to issue-related hashtags, *Figure 2* (see online supplement for hourly breakdown) reveals that pro-Remain hashtags were responsible for nearly 7% of tweets while pro-Leave ones generated twice as much traffic (13%).

During our data collection period, 176 accounts tweeted at an average rate of more than 50 times per day, generating 5% of the overall traffic. Out of these, 87 were actively tweeting during the spike in traffic around Conservative hashtags on November 19th. However, after running these accounts through the Botometer API, an open source tool trained to classify an account as bot or human based on activity patterns, language and social structure, only 3 accounts returned a higher than 50% chance of being managed by a bot while 5 had been either deleted or suspended. Due to ethical considerations, we did not analyse the content of those tweets or the provenance of these accounts. Hence, this information alone is insufficient to determine whether these prolific Twitter accounts were simply run by motivated voters or by outsiders tasked with denouncing or bolstering the Conservatives' digital tactics.

**Table 1 - TYPES OF POLITICAL NEWS AND INFORMATION SHARED OVER TWITTER (%)**
BETWEEN MIDNIGHT 12–19 NOVEMBER 2019

| Type of Source | Percentage |
| --- | --- |
| **Professional News Content** | **57.1** |
| Major News Brands | 33.0 |
| Local News | 6.8 |
| New Media and Start-Ups | 5.7 |
| Tabloid | 11.6 |
| **Professional Political Content** | **16.1** |
| Government | 7.4 |
| Experts | 1.3 |
| Political Party or Candidate | 7.4 |
| **Junk News & Propaganda** | **1.8** |
| Junk News Sites | 1.7 |
| Russian Propaganda Content | 0.1 |
| **Other Political Information** | **20.7** |
| Citizen, Civil Society and Civic Content | 11.8 |
| Video/Image Sharing and Content Subscriptions | 3.5 |
| Online Portals, Search Engines and Aggregators | 2.8 |
| Other | 2.6 |
| **Miscellaneous** | **4.1** |
| Social Media Platform | 1.5 |
| Other | 2.6 |

**Figure 1 - HOURLY TWITTER TRAFFIC AROUND ELECTION AND PARTY-RELATED HASHTAGS**
BETWEEN MIDNIGHT 12–19 NOVEMBER 2019

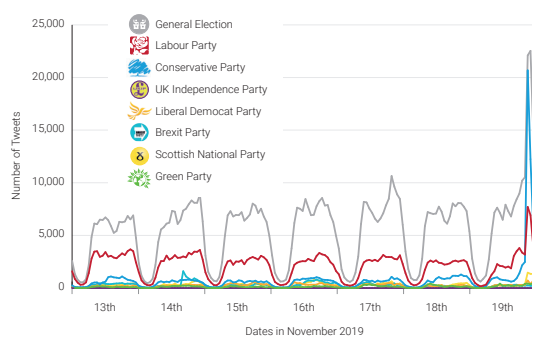

**Figure 2 - HOURLY TWITTER TRAFFIC AROUND ISSUE-RELATED HASHTAGS**
BETWEEN MIDNIGHT 12–19 NOVEMBER 2019

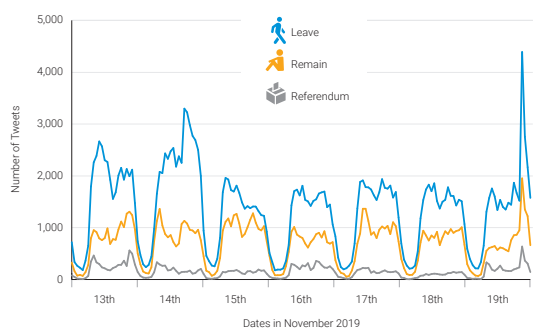

# Trends in User Interactions with Junk and Professional Content on Facebook

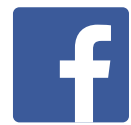

User interactions on Facebook are a useful way to gauge the popularity of and public engagement with different sources of political news and information. Using the CrowdTangle dashboard, our team measured the volumes of interactions (comments, likes, and shares) with content produced by the eighteen most popular sources of junk and professional news in our dataset during the first three weeks of the UK election campaign.

As *Table 2* shows, junk news sites were less prolific publishers than professional news ones over that time period, posting an average of 10 stories per day compared to 38 for major news organizations. This holds true when one excludes prolific US junk news outlets, such as *Breitbart*, from the analysis (see online supplement for full breakdown). Stories from *The Independent* alone, for example, were shared over 987,000 times between November 6th and November 27th — nearly 6 times more than stories from the top 3 non-US junk news outlets. Out of the 30 outlets classified as junk news by our team, 10 primarily focus on the US while selectively reporting on UK political events.

Despite the inclusion of *Breitbart*, which slightly skewed our averages, it is noteworthy that stories from the top junk outlets in our sample were still shared less widely overall (88,962 times per outlet) than those from major professional news outlets (137,619 times per outlet). Beyond shares alone, however, junk and professional media outlets almost overlap in terms of public engagement with their content (see *Table 2*). Between 6th and 27th November mainstream stories received an average of 820 interactions, including "likes" and "comments" while that number stood at 790 for lower quality and more polarizing content.

Junk news sites often disseminate fewer outputs and have a smaller readership than their professional counterparts, especially when compared to legacy media publishers. Nevertheless, their stories can elicit angrier and more outraged responses from Facebook users. Content posted by junk news outlets tends to be more visual, with stories being more often shared on Facebook in the form of photo posts than the average story from a professional news brand (see *Figure 4*). Not only that, but stories published by junk news outlets provoke more extreme or emotive reactions. In 40% of cases in our sample, junk posts triggered either an angry ("Angrys") or laughing ("Hahas") response on Facebook. This underscores the polarizing potential of news operations whose business models reward clickbait, viral content and outrage-mongering. In contrast, mainstream stories shared over Facebook during the first three weeks of the campaigns garnered more moderate and evenly distributed reactions from users.

**Figure 3 - USER REACTIONS TO JUNK AND PROFESSIONAL NEWS POSTS**
BETWEEN 6–27 NOVEMBER 2019

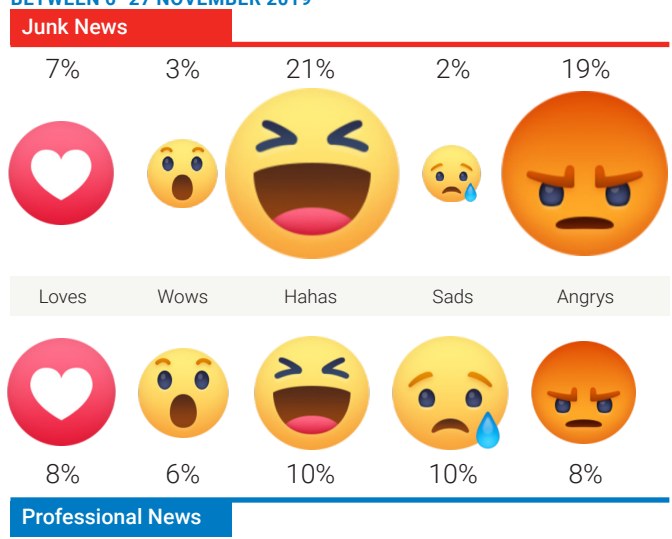

**Table 2 - AVERAGE PUBLIC ENGAGEMENT WITH TOP JUNK NEWS AND PROFESSIONAL NEWS OUTLETS**
BETWEEN 6–27 NOVEMBER 2019

|   | Total Posts | Interactions | Shares | Posts Per Day | Avg. Interactions/Post | Page Likes |
|---|---|---|---|---|---|---|
| **Junk News** | 201 | 656,851 | 88,962 | 9.6 | 790 | 279,852 |
| **Professional News** | 802 | 704,886 | 137,619 | 38.2 | 820 | 5,139,054 |

# Key Themes across Popular Junk News Stories

Having explored patterns of information sharing over both Facebook and Twitter, we then proceeded to isolate the most popular stories from junk news sites shared by UK Facebook pages on each day of our data collection. Rather than peddling entirely made-up facts, nearly every story in this sample spun reporting by more established outlets — often distorting or exaggerating the truth — serving ideological agendas in the process. Such stories were primarily shared by Facebook pages focused on nationalist themes, pro-Leave and pro-Remain campaigning groups, as well as the pages of known junk news outlets themselves.

A high-level analysis of the main themes of these stories reveals that most of them propagated anti-mainstream media narratives. Out of the 22 stories we analysed for this exercise, 8 explicitly referred to the "BBC", "the mainstream media" or other legacy newspapers and political journalists in their headlines and leads (see *Figure 5*). In 75% of cases, this was coupled with accusations of wrongdoing, bias or lying. Several viral stories, for instance, accused the BBC of "covering up" and "siding" with Boris Johnson after it was revealed the public broadcaster had edited out audience laughter at the Prime Minister's expense in a video clip shared in the wake of the Question Time leader debate. Other stories similarly indicted the mainstream media for obscuring the truth or suppressing information to favour one side or another of the political aisle.

In contrast, fewer stories focused on the parties' agendas and proposed policies (4 out of 22). There were two notable exceptions to this trend: the story that generated the most buzz celebrated Boris Johnson's proposal to toughen up Britain's immigration laws, while the most viral native video condemned Conservative home secretary Priti Patel for absolving the government of responsibility for mounting poverty in the UK. Only two posts directly referred to Labour policies, one of which criticized the BBC for framing Jeremy Corbyn's proposal for universal broadband as "communism."

**Figure 4 - PROPORTION OF POSTS CONTAINING A PHOTO OR VIDEO**
BETWEEN 6–27 NOVEMBER 2019

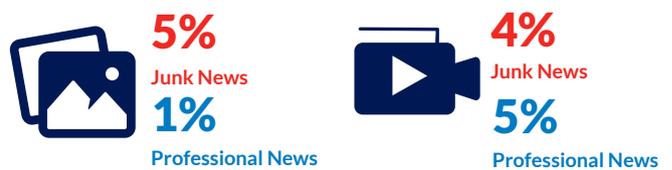

5% Junk News
1% Professional News

4% Junk News
5% Professional News

**Figure 5 - TOP STORIES FROM MOST SHARED JUNK NEWS SITES**
BY TOTAL INTERACTIONS BETWEEN 6–27 NOVEMBER 2019

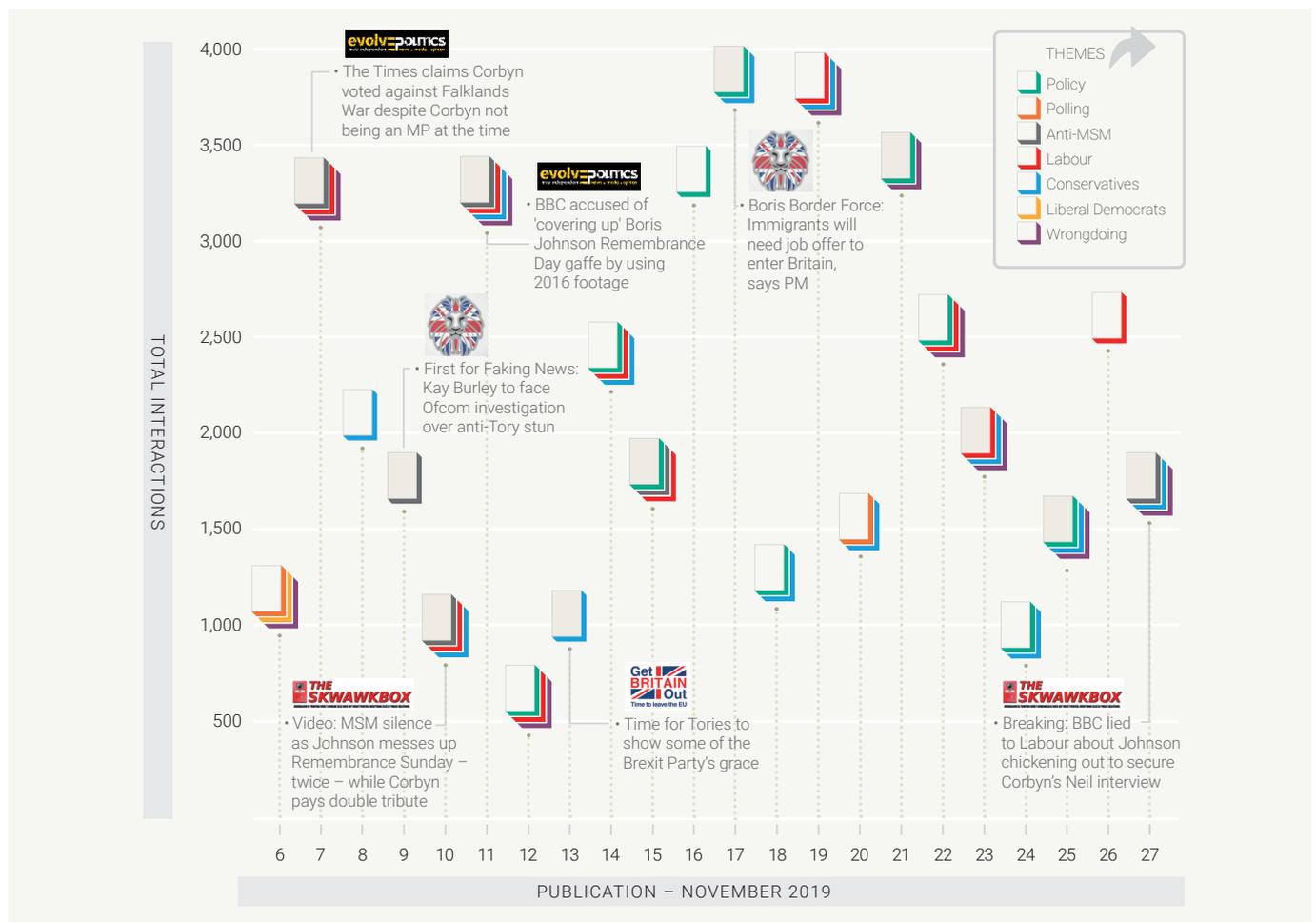

# Conclusion

Our research shows that sharing divisive, conspiratorial and low-quality information over platforms like Facebook and Twitter are common tactics to manipulate public opinions. Yet, in Western European democracies, this practice seems to be on the decline. In 2017, our team had found that junk news and traffic manipulation were prominent during the UK General Election campaign. During the most recent European Parliamentary elections, however, we found that less than 4% of the sources circulating on Twitter in seven language spheres were junk news, with users sharing much higher proportions of links to professional news sources overall.[13]

Echoing this trend, in this memo we find that **(1)** less than 2% of the links shared over Twitter during our data collection redirected to junk news sites, with users sharing higher proportions of links to professional news sources overall; **(2)** on Facebook, stories from junk news outlets were shared by far less people overall, though they triggered more extreme responses from users; **(3)** the most viral junk news stories in our dataset aimed to discredit the mainstream news media, with few covering party policies or political agendas.

Taken together, these findings indicate that regular users following the political conversation on Twitter and Facebook through hashtags before the vote were mostly sharing links to high-quality news, including high volumes of content produced by independent citizen, civic groups and civil society organizations, and were less likely to be exposed to polarizing content.

## About the Project

The Computational Propaganda Project (COMPROP), which is based at the Oxford Internet Institute, University of Oxford, is an interdisciplinary team of social and information scientists researching how political actors manipulate public opinion over social networks. This work includes analysing how the interaction of algorithms, automation, politics, and social media amplifies or represses political content, disinformation, hate speech, and junk news. Fact sheets integrate important trends identified during analyses of current events with basic data visualizations, and although they reflect methodological experience and considered analysis, they have not been peer reviewed. Working papers present deeper analysis and extended arguments that have been collegially reviewed and engage with public issues. COMPROP's articles, book chapters, and books are significant manuscripts that have been through peer review and formally published.

## Acknowledgments

The authors gratefully acknowledge the support of the European Research Council for the project 'Computational Propaganda: Investigating the Impact of Algorithms and Bots on Political Discourse in Europe', Proposal 648311, 2015–2020, Philip N. Howard, Principal Investigator. Project activities were approved by the University of Oxford's Research Ethics Committee, CUREC OII C1A 15-044. We are also grateful to Omidyar Network, Adessium and Civitates, for supporting our Election Observatory and our research in Europe. Any opinions, findings, conclusions or recommendations expressed in this material are those of the authors and do not necessarily reflect the views of the University of Oxford or our funders. We are grateful to Zeeshan Ali, Dragos Gorduza and Dr. Vidya Narayanan for their contributions to this fact sheet.